\documentclass[11pt]{article}
\usepackage{times}
\usepackage{sidecap}
\usepackage{geometry}
\usepackage{floatrow}
\usepackage{float}
\usepackage{hyperref}
\usepackage{graphics,graphicx}
\usepackage{subcaption}
\usepackage{epsf,wrapfig}

\usepackage{natbib,bm}

\usepackage{paralist}
\let\olditem\item
\renewenvironment{thebibliography}[1]{%
  \section*{\refname}
  \let\par\relax
  \renewcommand{\item}[1][]{\olditem[\textbullet]}%
  \inparaenum}{\endinparaenum}

\newcounter{aff}

\renewenvironment{thebibliography}[1]{%
\textsc{\textbf{References:}} 
\let\par\relax%
\inparaitem[{[}1{]}]}{\endinparaitem}

\DeclareGraphicsExtensions{.jpg,.pdf,.png,.eps}

\usepackage{caption}
\usepackage{xcolor}

\usepackage{tcolorbox}

\bibpunct{(}{)}{;}{a}{}{,}

\usepackage{soul}

\definecolor{darkcyan}{rgb}{0.0, 0.0, 0.7}

\definecolor{boxcolor}{RGB}{193, 229, 255}
\definecolor{figboxcolor}{RGB}{193, 229, 255}
\definecolor{boxcolorgray}{RGB}{227, 227, 227}
\definecolor{darkcyan}{rgb}{0.0, 0.0, 0.7}

\sethlcolor{boxcolorgray}

\def\gtrsim{\lower 2pt \hbox{$\, \buildrel {\scriptstyle >}\over
{\scriptstyle \sim}\,$}}
\def\lesssim{\lower 2pt \hbox{$\, \buildrel {\scriptstyle <}\over
{\scriptstyle \sim}\,$}}

\def\approxlt{\lower.2em\hbox{$\buildrel < \over \sim$}}
\def\approxgt{\lower.2em\hbox{$\buildrel > \over \sim$}}

\def\be{\begin{equation}}
\def\ee{\end{equation}}
\def\bea{\begin{eqnarray*}}
\def\eea{\end{eqnarray*}}

\def\jref@jnl#1{{\rm#1}}
\def\aj{\jref@jnl{AJ}}			
\def\araa{\jref@jnl{ARA\&A}}		
\def\apj{\jref@jnl{ApJ}}			
\def\apjl{\jref@jnl{ApJ}}		
\def\apjs{\jref@jnl{ApJS}}		
\def\ao{\jref@jnl{Appl.~Opt.}}		
\def\apss{\jref@jnl{Ap\&SS}}		
\def\aap{\jref@jnl{A\&A}}		
\def\aapr{\jref@jnl{A\&A~Rev.}}		
\def\aaps{\jref@jnl{A\&AS}}		
\def\azh{\jref@jnl{AZh}}			
\def\baas{\jref@jnl{BAAS}}		
\def\jrasc{\jref@jnl{JRASC}}		
\def\memras{\jref@jnl{MmRAS}}		
\def\mnras{\jref@jnl{MNRAS}}		
\def\pra{\jref@jnl{Phys.~Rev.~A}}	
\def\prb{\jref@jnl{Phys.~Rev.~B}}	
\def\prc{\jref@jnl{Phys.~Rev.~C}}	
\def\prd{\jref@jnl{Phys.~Rev.~D}}	
\def\pre{\jref@jnl{Phys.~Rev.~E}}	
\def\prl{\jref@jnl{Phys.~Rev.~Lett.}}	
\def\pasp{\jref@jnl{PASP}}		
\def\pasj{\jref@jnl{PASJ}}		
\def\qjras{\jref@jnl{QJRAS}}		
\def\skytel{\jref@jnl{S\&T}}		
\def\solphys{\jref@jnl{Sol.~Phys.}}	
\def\sovast{\jref@jnl{Soviet~Ast.}}	
\def\ssr{\jref@jnl{Space~Sci.~Rev.}}	
\def\zap{\jref@jnl{ZAp}}			
\def\nat{\jref@jnl{Nature}}		
\def\iaucirc{\jref@jnl{IAU~Circ.}}
\def\aplett{\jref@jnl{Astrophys.~Lett.}}
\def\apspr{\jref@jnl{Astrophys.~Space~Phys.~Res.}}
\def\bain{\jref@jnl{Bull.~Astron.~Inst.~Netherlands}}
\def\fcp{\jref@jnl{Fund.~Cosmic~Phys.}}
\def\gca{\jref@jnl{Geochim.~Cosmochim.~Acta}}
\def\grl{\jref@jnl{Geophys.~Res.~Lett.}}
\def\jcp{\jref@jnl{J.~Chem.~Phys.}}	
\def\jgr{\jref@jnl{J.~Geophys.~Res.}}	
\def\jqsrt{\jref@jnl{J.~Quant.~Spec.~Radiat.~Transf.}}
\def\memsai{\jref@jnl{Mem.~Soc.~Astron.~Italiana}}
\def\nphysa{\jref@jnl{Nucl.~Phys.~A}}
\def\physrep{\jref@jnl{Phys.~Rep.}}
\def\physscr{\jref@jnl{Phys.~Scr}}
\def\planss{\jref@jnl{Planet.~Space~Sci.}}
\def\procspie{\jref@jnl{Proc.~SPIE}}

\def\farcs{\hbox{$.\!\!^{\prime\prime}$}}

\def\ion#1#2{#1$\;${\small\rm\@Roman{#2}}\relax}

\newcount\lecurrentfam
\def\LaTeX{\lecurrentfam=\the\fam \leavevmode L\raise.42ex
\hbox{$\fam\lecurrentfam\scriptstyle\kern-.3em A$}\kern-.15em\TeX}
\def\sizrpt{
(\fontname\the\font): em=\the\fontdimen6\font, ex=\the\fontdimen5\font
\typeout{
(\fontname\the\font): em=\the\fontdimen6\font, ex=\the\fontdimen5\font
}}
\definecolor{dark}{rgb}{ 0.34, 0.09, 0.49}

\usepackage{fancyhdr,url}
\usepackage{transparent}
\usepackage{tablefootnote}
\usepackage{mathpazo}

\DeclareRobustCommand{\ion}[2]{\textup{#1\,\textsc{\lowercase{#2}}}}
\newcommand*\element[1][]{%
  \def\aa@element@tr{#1}%
  \aa@element
}

\begin{document}

\noindent
\begin{raggedright}
{\Huge\bfseries AGN Feeding and Feedback: Resolving the Torus \\[0.5em]
\Large How does the torus feed the accretion disk? \\[0.3em]
How does feeding/feedback work in local quasars?
}\end{raggedright}\\

\noindent
M. Pereira-Santaella$^1$, 
F. Combes$^2$, 
A. Alonso-Herrero$^3$, 
S. Garc\'ia-Burillo$^4$,
A. Audibert$^5$,
E. Bertola$^6$,
M. Bischetti$^7$,
R. Decarli$^8$,
I. Garc\'ia-Bernete$^3$,
I. Lamperti$^9$,
I. Ruffa$^{6,10}$,
F. Salvestrini$^{7}$,
H. R. Stacey$^{11}$,
M. V. Zanchettin$^{12}$,
V. Casasola$^{10}$,
D. Espada$^{13,14}$,
D.\ Nguyen$^{15}$,
M. Pandey-Pommier$^{16}$,
S. Thater$^{17}$

\vspace{1em}
{
\footnotesize{
\noindent $^1$ Instituto de F\'isica Fundamental, CSIC, Calle Serrano 123, 28006 Madrid, Spain \\
$^2$ LUX, Observatoire de Paris, Coll\`ege de France, PSL University
CNRS, Sorbonne University, Paris, France \\
$^3$ Centro de Astrobiolog\'ia (CAB), CSIC-INTA, Camino Bajo del Castillo s/n, 28692 Villanueva de la Ca\~nada, Madrid, Spain \\
$^4$ Observatorio Astron\'omico Nacional (OAN-IGN)-Observatorio de Madrid, Alfonso XII, 3, 28014 Madrid, Spain \\
$^5$ Instituto de Astrof\'isica de Canarias, Calle V\'ia L\'actea, s/n, 38205 La Laguna, Tenerife, Spain \\
$^6$ INAF-OAA, Osservatorio Astrofisico di Arcetri, largo E. Fermi 5, 50127 Firenze, Italy \\
$^7$ Dipartimento di Fisica, Universit\'a di Trieste, Sezione di Astronomia, Via G.B. Tiepolo 11, 34131 Trieste, Italy \\
$^8$ INAF, Osservatorio di Astrofisica e Scienza dello Spazio di Bologna, via Gobetti 93/3, 40129 Bologna, Italy \\
$^9$ Centro de Astrobiolog\'ia (CAB), CSIC-INTA, Ctra de Ajalvir, km 4, 28850 Torrej\'on de Ardoz, Madrid, Spain \\
$^{10}$ INAF - Istituto di Radioastronomia, via P. Gobetti 101, 40129 Bologna, Italy \\
$^{11}$ European Southern Observatory (ESO), Karl-Schwarzschild Str. 2, 85748 Garching, Germany \\
$^{12}$ INAF -- Istituto di Astrofisica e Planetologia Spaziali, Via Fosso del Cavaliere 100, 00133 Rome, Italy  \\
$^{13}$ Departamento de F\'{i}sica Te\'{o}rica y del Cosmos, Campus de Fuentenueva, Edificio Mecenas, Universidad de Granada, 18071 Granada, Spain \\
$^{14}$ Instituto Carlos I de F\'{i}sica Te\'{o}rica y Computacional, Facultad de Ciencias, 18071 Granada, Spain \\
$^{15}$ Department of Astronomy, University of Michigan, 1085 South University Avenue, Ann Arbor, MI 48109, USA \\
$^{16}$ Pole Scientific, University Catholic of Lyon- University of Lyon, 10 place des Archives, 69002 Lyon, France \\
$^{17}$ Department of Astrophysics, University of Vienna, Türkenschanzstraße 17, 1180 Vienna, Austria \\

}
}
\begin{tcolorbox}[boxsep=3pt,left=0pt,right=0pt,top=1pt,bottom=1pt,colframe=boxcolor,colback=boxcolor]
\justify
\textbf{Abstract:} 
Active Galactic Nuclei (AGN) play a key role in galaxy evolution by regulating star formation within their host galaxies.
It is therefore crucial to understand how the central supermassive black holes (SMBH) are fueled by material from the galactic disk, and how AGN feedback impacts the surrounding gas.
Most information so far comes from very nearby active galaxies ($z < 0.008$; $D$$<$35\,Mpc), providing the best spatial resolution and revealing randomly oriented {\it dusty molecular tori} that fuel SMBHs.
A facility such as ALMA2040, offering significantly higher sensitivity and better angular resolution than current ALMA, would enable progress in two fundamental areas that remain beyond the capabilities of all existing and planned observatories: 
(i) probing the dynamics of the material feeding the AGN down to the accretion disk in nearby galaxies; 
and (ii) spatially resolving the tori and circum-nuclear regions of more energetic and cosmologically significant AGN, such as local quasars ($z$$\sim$0.1 and $L_{\rm bol}$$>$10$^{45}$\,erg\,s$^{-1}$).
\end{tcolorbox}

\newpage{}

\vspace{-0.8cm}
\section{Scientific context and open questions}
\vspace{-0.2cm}
Despite their very different spatial scales, the evolution of galaxies and the growth of the super-massive black holes (SMBH) in their centers are closely connected, as revealed by the scaling relations between the mass of the SMBH and stellar mass of the bulge \citep{Kormendy2013ARA&A..51..511K}.
It is now well established that SMBH can regulate the formation of stars in their host galaxies through radiative feedback, radio jets, and massive outflows driven by the large amounts of energy released during accretion in the active galactic nucleus (AGN) phase \citep{Harrison2024Galax..12...17H}.
Indeed, AGN feedback is crucial in galaxy formation models and simulations to reproduce the observed properties of massive galaxies \citep{Springel2005MNRAS.361..776S, Vogelsberger2020}.

The central element of the Unified Model for AGN is the torus of dust and gas around the SMBH
\citep[e.g.,][]{Krolik1988ApJ...329..702K, Antonucci1993ARA&A..31..473A}.
This key structure not only explains the observational distinction between type 1 AGN, where both broad and narrow line regions (BLR and NLR) are visible, and type 2 AGN, where the torus obscures the BLR, but, more importantly, it plays a critical role in the final stages of AGN fueling as the main reservoir of material around the SMBH.

ALMA has recently made a breakthrough in our understanding of how AGN are fueled and how they regulate star formation through feedback. Circum-nuclear molecular disks and molecular tori have been discovered in many 
nearby Seyfert AGN of intermediate luminosity \linebreak ($L_{\rm bol}$=10$^{42-44}$\,erg\,s$^{-1}$) on $\sim$10\,pc scales (e.g., \citealt{GarciaBurillo2016ApJ...823L..12G,
GarciaBurillo2021A&A...652A..98G, 
AlonsoHerrero2018ApJ...859..144A, 
AlonsoHerrero2019A&A...628A..65A, 
AlonsoHerrero2023A&A...675A..88A, 
Combes2019A&A...623A..79C, 
Combes2026A&A...705A.124C,
Audibert2019A&A...632A..33A,
Gallimore2016ApJ...829L...7G, 
Imanishi2018ApJ...853L..25I})
revealing through dust continuum and several molecular lines (CO, HCO$^+$, and HCN) this key element of the AGN unification paradigm. These tori appear kinematically decoupled from their host, in some cases exhibiting counter-rotation \citep{Impellizzeri2019ApJ...884L..28I, Imanishi2020ApJ...902...99I}. Our present view of molecular tori is more complex than the iconic model brought by optical, X-ray, and radio observations \citep{Urry1995PASP..107..803U},
as observations increasingly reveal a complex, clumpy, and dynamically diverse tori and circum-nuclear medium. Moreover, below the 10\,pc scale, our models are still unconstrained, with only some elements being known from rare VLBI observations, reverberation mapping, or near-IR interferometry (e.g., \citealt{Greenhill1995A&A...304...21G, Greenhill1996ApJ...472L..21G, Uttley2014A&ARv..22...72U, Gravity2018Natur.563..657G}).

\begin{figure*}[t]
    \centering
    \includegraphics[width=\textwidth]{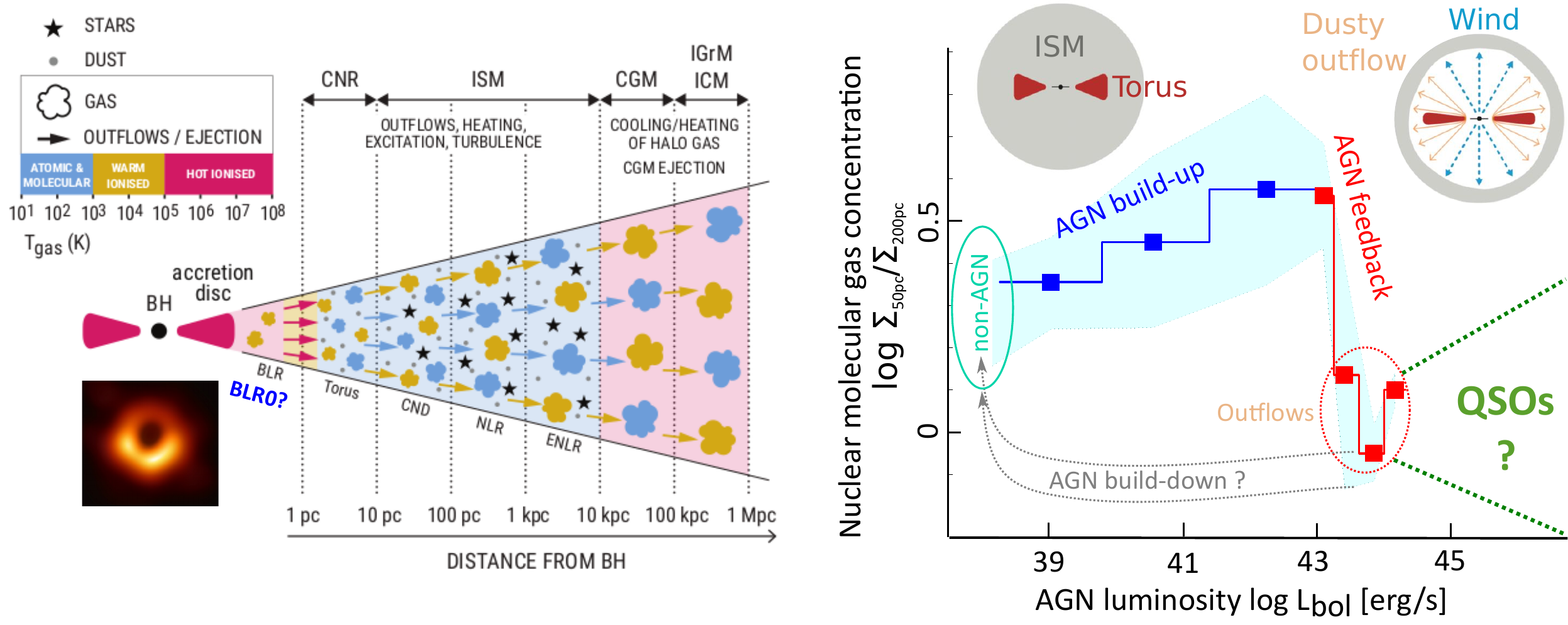}
    \caption{{\bf Left. }{\it Diagram showing the orders-of-magnitude range in spatial scales, from nuclear to galaxy cluster, and gas phases, from hot-ionized to molecular,
    of AGN feedback.
    Adapted from \cite{Harrison2024Galax..12...17H,EHT2019ApJ...875L...1E}.} {\bf Right. } {\it Nuclear molecular gas concentration as a function of AGN luminosity for nearby Seyfert galaxies \citep{GarciaBurillo2024A&A...689A.347G}. The turnover at $L_{\rm bol}$$\sim$ 10$^{42.5}$\,erg\,s$^{-1}$ ($L_{2-10\,{\rm keV}}$$\sim$ 10$^{41.5}$\,erg\,s$^{-1}$), divides the sample into the build-up and feedback branches. The left inset represents the limited effect of AGN feedback on the torus and ISM during the build-up branch, while the right inset shows the imprint left by AGN feedback on the distribution of molecular gas in the torus and its surroundings through the launching of ionized winds and dusty outflows.
    The regime corresponding to higher $L_{\rm bol}$ sources, QSOs, remains to be explored. Adapted from \citep{GarciaBurillo2024A&A...689A.347G,GarciaBurillo2021A&A...652A..98G}.}}
    \label{fig:scales}
\end{figure*}

At the same time, AGN-driven winds are expected to expel part of the torus and drive massive multi-phase outflows that can extend up to $\sim$kpc scales, impacting the interstellar medium of the host galaxy and producing cavities devoid of molecular gas (e.g., \citealt{MercedesFeliz2023,MercedesFeliz2026}).
These cyclical and simultaneous feeding\slash feedback processes are fundamental to our understanding of SMBH growth and the AGN effects on their host galaxy (Fig.~\ref{fig:scales}).

In this white paper, we focus on two key aspects of these processes that remain largely unexplored due to the limited angular resolution and sensitivity of current observatories:
\vspace{-0.1cm}
\begin{itemize}
\item {\bf How does the material in the torus feed the AGN?}
To reach the SMBH, the molecular gas in the torus must lose angular momentum, pass through the dust sublimation radius, a potential molecular dust-free neutral BLR, the atomic BLR, and ultimately be funnelled into the accretion disk. %
\vspace{-0.1cm}
\item {\bf How does feeding\slash feedback operate in the most luminous AGN?}
Throughout most of the Universe history, high-luminosity AGN ($L_{\rm bol}$$>$10$^{45-47}$\,erg\,s$^{-1}$), such as quasars (QSOs), have dominated AGN luminosity.
However, due to their greater distances ($D>$250\,Mpc), the tori and the surroundings fueling these rapidly accreting SMBHs remain unresolved.
The torus geometry, winds, and outflows depend on the AGN luminosity (Fig.~\ref{fig:scales} right), making it essential to resolve the inner regions of QSOs to understand feedback across cosmic times and physical scales.

\end{itemize}

\section{Science cases for ALMA2040}

\subsection{Neutral molecular BLR and resolved torus}

Around the SMBH at the center of AGN, are expected to exist accretion disks of ionized gas that are fueled by a BLR. The inner part of this accretion disk has been mapped by the Event Horizon Telescope (EHT) around M87* and Sgr~A* in 2019 and 2022, respectively \citep{EHT2019ApJ...875L...1E, EHT2022ApJ...930L..12E}. 
The BLR may extend up to 1000 Schwarzschild radii (or 10$^{-2}$\,pc, for $M_{\rm BH}$=10$^8$\,$M_\odot$) and contain gas gravitationally bound rotating at velocities of more than a few 1000 kms$^{-1}$. However, between the molecular torus and the accretion disk, it is not yet clear in what form the matter exists. Dust can survive up to temperatures of the order of 1500\,K, and inside the sublimation radius, there might exist neutral molecular gas, which progressively gains high rotational velocity. Close to the SMBH, this could represent a BLR$^0$ region, i.e., a broad line region of dustless neutral gas. In this region water masers have been observed, at radii less than 1\,pc (e.g., \citealt{Lo2005ARA&A..43..625L}).

\begin{figure*}[t]
    \centering
    \includegraphics[width=0.75\textwidth]{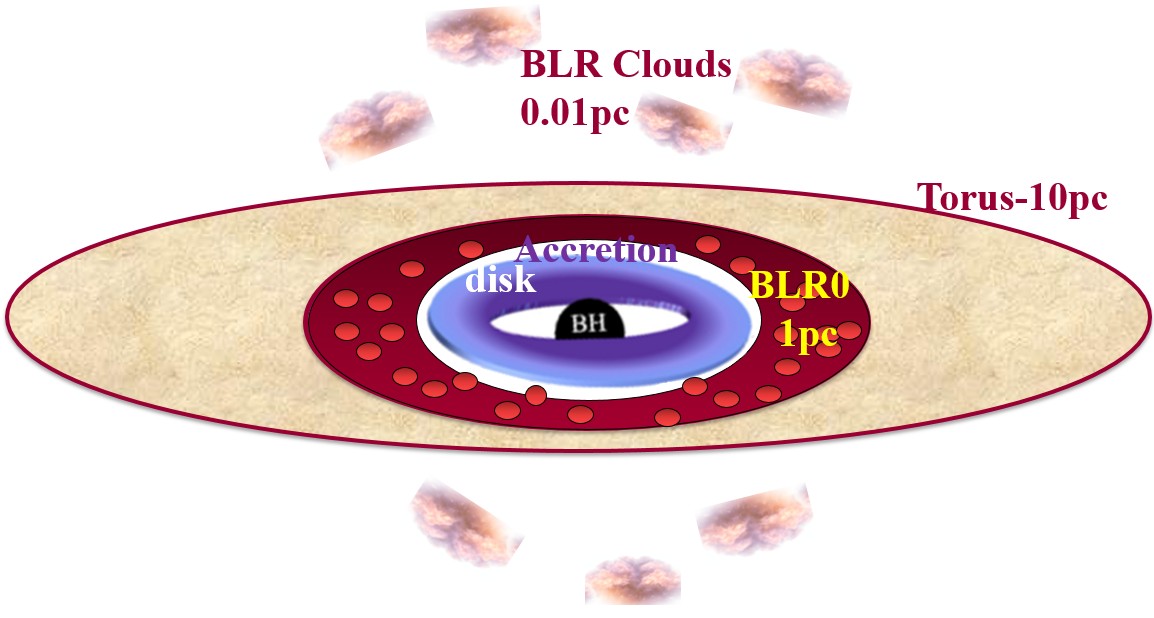}
    \caption{\it Schematic diagram showing the disposition of the circum-nuclear medium around the SMBH of a typical AGN, where M$_{BH}$= 10$^8$ M$_\odot$. Around the center, the accretion disk (in blue), of size 1000 to 10000 AU;
    surrounded by the ionized BLR clouds of size up to 0.01pc; then a thin and dense neutral disk without dust (in red), the BLR$^0$ of typical size 1pc. At 10pc scale, the molecular and dusty torus. The components are not to scale. The tracers are ground-state CO for the torus, CO $v$=1 for the proposed BLR$^0$, and \ion{O}{i}, \ion{Ca}{ii}, and hydrogen lines for the optical components.}
    \label{fig:blr0}
\end{figure*}

According to a recent proposal by \citet{Thi2024A&A...688L..20T}, there might exist a significant mass ($\sim$10$^5$\,$M_\odot$) of neutral atomic and/or molecular gas, in hot, dense, and dustless regions, at high velocities around the SMBH. This neutral phase might reside in a relatively thin disk (BLR$^0$), while the ionized BLR rotates around, in a thicker component, fragmented in BLR clouds (see the schematic drawing in Fig.\,\ref{fig:blr0}). While the ionized BLR emits UV/optical lines, 10$^3$--10$^4$\,km\,s$^{-1}$ wide, and has a mass 10--100\,$M_\odot$ at a typical radius of $\sim$10$^{-2}-10^{-3}$\,pc for a typical SMBH mass of $\sim$10$^8$\,$M_\odot$, the neutral BLR$^0$ could have a mass of up to 10$^5$\,$M_\odot$, so containing the bulk of the matter.

ALMA2040 could detect this BLR$^0$, even if it has not sufficient spatial resolution to make a spatial map. The detection would be through the wide velocity width of molecular lines, analogous to the detection of BLRs in the UV/optical using broad atomic lines.

\begin{figure*}[t]
\vspace{-0.4cm}
    \centering
    \includegraphics[width=0.65\textwidth]{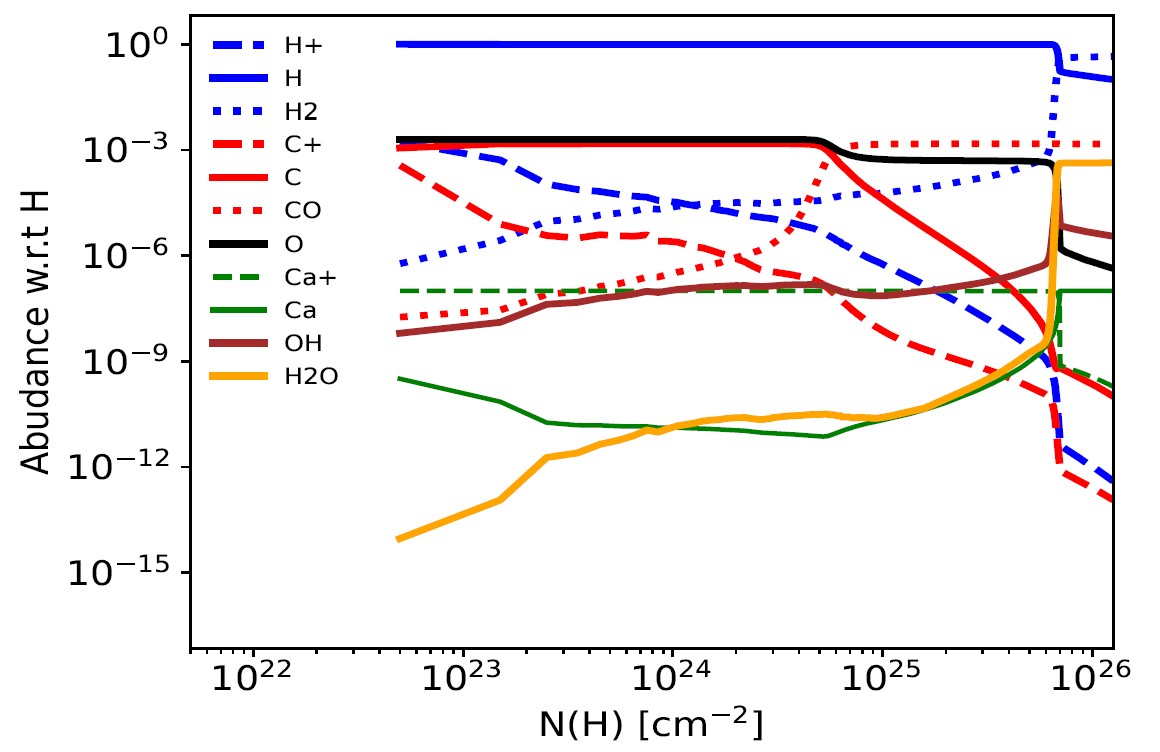}
    \caption{\it Prediction of species abundances, from \textsc{cloudy}, with log($U$)= --5, $U$ being the dimensionless ionization parameter \citep{Thi2024A&A...688L..20T}.}
    \label{fig:blr0_model}
    \vspace{-0.3cm}
\end{figure*}

The hot and dense molecular gas within the BLR$^0$ will excite the first vibrational state of the CO molecule. Consequently, this region is the only environment near the AGN where these vibrationally excited CO lines, with frequencies different from $v$=0 lines, are expected to arise.
Already in the changing look AGN NGC\,1566, 
\citet{Ochmann2024A&A...686A..17O} reveal a neutral component in a double-peaked BLR spectrum of \ion{O}{i} and \ion{Ca}{ii} triplet, interpreted as an eccentric accretion disk with Balmer lines at higher altitude. Calculations with a \textsc{cloudy} model, with number densities $\sim$10$^{10}$--10$^{14}$\,cm$^{-3}$ and ionization parameter log($U$) = $-$5, reveals that the molecular gas will dominate, with column densities as high as $N_{\rm H_2}$ $\sim$10$^{25-26}$\,cm$^{-2}$ (Fig.\ref{fig:blr0_model}). 

Several lines could be targeted, from the pure rotational ladder of CO, and also \ion{C}{i} and H$_2$O. Very close to the AGN, the strong radiation may create an X-ray dominated region (XDR) where the CO molecules might be dissociated, thus increasing the abundance of atomic carbon \linebreak (e.g.,\citealt{Maloney1996ApJ...466..561M, Bisbas2015ApJ...803...37B, Bisbas2017ApJ...839...90B}). \ion{C}{i} will better trace the neutral atomic gas in regions reached by cosmic-rays, and especially close to synchrotron jets, where the \ion{C}{i}(1--0) line at 492\,GHz can be 20--30 times brighter than the CO $v$=0 (3--2) line \citep{Papadopoulos2018MNRAS.478.1716P}.
However, for the molecular BLR$^0$, the best lines would be the vibrationally excited CO $v$=1 lines as they are exclusively produced in this region.
Current ALMA sensitivity is not sufficient for this detection in realistic integration times (see Table~\ref{tab:co_vib}).
Therefore, increasing the collecting area as proposed by ALMA2040 is essential. Additionally, the broad line width ($\sim$2000\,km\,s$^{-1}$) requires at least the increased bandwidth provided by the WSU update of ALMA in order to properly subtract the continuum emission.

The BLR$^0$ will be identified through its broad line wings, which distinguish it from the torus emission. It will still appear unresolved as a point source. However, high angular resolution observations are needed to isolate the central emission from the molecular torus and avoid its dilution and confusion. 
To be distinguished clearly from a possible outflow, the CO $v$=1 lines 
will be observed simultaneously with the ground state CO $v$=0 lines and other tracers of XDR and cosmic-ray dominated regions (e.g., [\ion{C}{i}]), thanks to the broad bandwidth of ALMA2040 ($>$512\,GHz). The frequencies of these CO lines are listed in Table\,\ref{tab:co_vib}.

In particular, the extended ALMA2040 configuration will provide a resolution better than 9\,mas at $\gtrsim$340\,GHz (0.5\,pc at 12\,Mpc for a typical nearby AGN) needed to isolate from the host ISM the torus and BLR$^0$ emissions for the CO $v$=0 (3--2) and $v$=1 (3$-$2) lines.
Given the large velocity gradients, and the clumpy nature of the gas in the torus, as well as in BLR$^0$, we do not expect to lose much emission with the Maximum Recoverable Scale MRS of 0\farcs2. We have estimated the peak flux expected from the BLR$^0$ in the CO $v$=0 (3$-$2) in Table\,\ref{tab:co_vib}. Previous ALMA observations of nearby AGN \citep{Combes2019A&A...623A..79C} have shown a molecular torus of $M$(H$_2$) = 10$^7$\,$M_\odot$, with a total of 40\,Jy\,km\,s$^{-1}$, and a full width half maximum (FWHM) of 200\,km\,s$^{-1}$, corresponding to 200\,mJy\,beam$^{-1}$. For a BLR$^0$ region, with $M$(H$_2$) = 10$^5$\,$M_\odot$, with a FWHM of 2000\,km\,s$^{-1}$, we expect a flux of 200\,$\mu$Jy\,beam$^{-1}$. The one hour on-source sensitivity of ALMA2040, \hbox{1$\sigma$ = 23\,$\mu$Jy\,beam$^{-1}$} for 100\,km\,s$^{-1}$ channels, will be sufficient for the detection at a 10$\sigma$ level of the $v$=0 line. This rough estimation does not take into account the different excitation between torus and BLR$^0$, which might be favorable to the BLR$^0$, with higher number and column densities.
Actually, the first vibrational level of CO is at $\sim$3100\,K above the ground level
and has been detected in several AGB stars in the warm layer between the stellar surface and the wind-starting region \citep{Khouri2016MNRAS.463L..74K}.
To estimate the intensity of the $v$=1 lines, we have assumed a kinetic temperature of 3000\,K, thus the population is lower than in $v$=0 by a factor $\sim$$\exp(-3100/T)$. However, since the CO $v$=0 (3-2) line is strongly optically thick ($\tau >$100), with the high column density mentioned above, the $v$=1 line will also be optically thick and therefore will be detectable, although the fluxes are quite uncertain. From these estimates, the required on-source integration time to detect the $v$=1 line at a 10$\sigma$ level in 300\,km\,s$^{-1}$ channels is $\sim$3.8\,h (see Table~\ref{tab:co_vib}).

Simultaneously, it will be also possible to sample the full continuum spectral energy distribution of the dusty torus, given the superb one hour on-source sensitivity of 0.2--2.3\,$\mu$Jy\,beam$^{-1}$ for 64\,GHz bins between 40 and 450\,GHz over the full $>$512\,GHz ALMA2040 bandwidth. Some of the continuum could be resolved out at high resolution, but we expect a highly clumpy character. 

Beyond mapping the feeding flow, resolving the torus and BLR$^0$ delivers a dynamical measurement of the SMBH mass from the same data. At the $\sim0.5$~pc spatial resolution delivered by ALMA2040 at $\sim$ 350\,GHz (9\,mas at 12\,Mpc), these observations can resolve the sphere of influence (e.g., $R_{\rm soi}\sim1$~pc for a $10^{7}\,M_\odot$ black hole for a galaxy with $\sigma\sim200$~km\,s$^{-1}$), so every torus target can simultaneously constrain $M_{\rm BH}$ from the gas kinematics. If the BLR$^0$ is a thin, rotationally supported disk recognized through its broad line wings, its velocity field provides a dust-free, molecular SMBH mass. Such measurements would anchor the $M_{\rm BH}$-$\sigma$ relation directly in active galaxies and complement direct dynamical SMBH masses (Decarli et al. 2026).

\begin{table}[h]
    \centering
    \begin{tabular}{lccccccccc}
    \hline
    \hline
    & $v$=1 & S ($\mu$Jy\,beam$^{-1}$) & $v$=0  & S ($\mu$Jy\,beam$^{-1}$) & Current ALMA & ALMA2040 \\ %
    & (GHz) & & (GHz) & &  &  \\  %
    \hline
CO(3$-$2) & 342.6476  & 70   &  345.7960  &  200 & 11.4\,days & 3.8\,h\\ %
\hline
    \end{tabular}
    \caption{Estimation of the expected line peak fluxes of the rotational lines of the first vibrationally excited CO levels for an AGN at 12\,Mpc with a BLR$^0$ region of $M$(H$_2$) = 10$^5$\,$M_\odot$. The last two columns compare the on-source integration time needed to detect the $v$=1 line with a signal to noise ratio of $\sim$10 in 300\,km\,s$^{-1}$ channels with current ALMA and ALMA2040. Frequencies from \citep{Gendriesch2009A&A...497..927G}.}
    \label{tab:co_vib}
\end{table}

\begin{figure*}[t]
    \centering
    \hfill
    \includegraphics[width=0.29\textwidth]{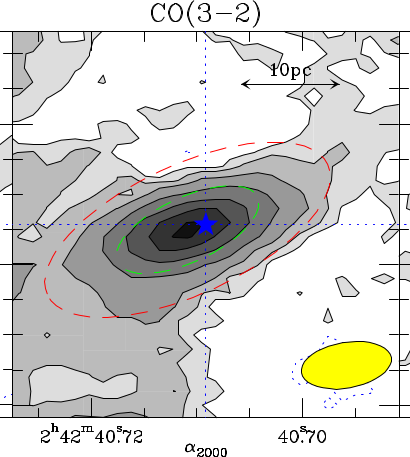}
    \hfill
    \includegraphics[width=0.50\textwidth]{Thi-CLOUDY.jpg}
    \hfill
    \caption{\it {\bf Left. }Molecular torus of NGC~1068 only marginally resolved by ALMA. The beam FWHM is the yellow ellipse \cite{GarciaBurillo2019A&A...632A..61G}. {\bf Right. }Abundance for various species as a function of the column density.  Model with n$_H$ = 10$^{14}$\,cm$^{-3}$, log($U$) = $-5$, and $\Delta$V = 250\,km\,s$^{-1}$.
    Molecular gas dominates at high $N_{\rm H}$ \cite{Thi2024A&A...688L..20T}.}
    \label{fig:torus_thi}
\end{figure*}

\begin{figure*}[t]
    \centering
    \hfill
    \includegraphics[width=0.49\textwidth]{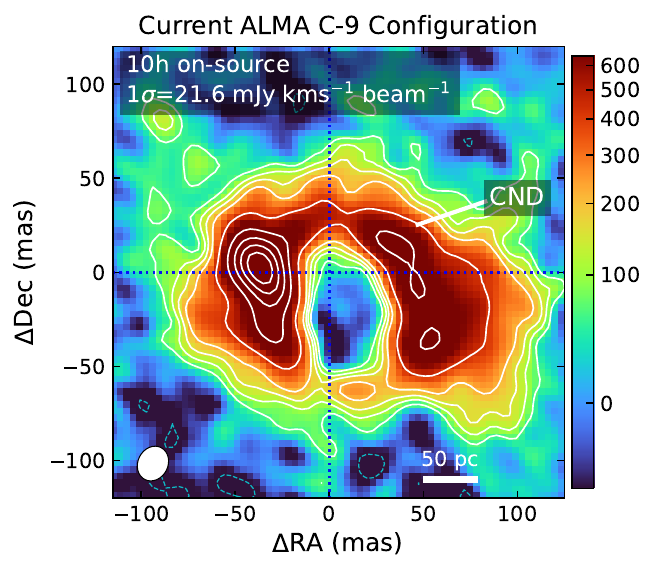}
    \hfill
    \includegraphics[width=0.49\textwidth]{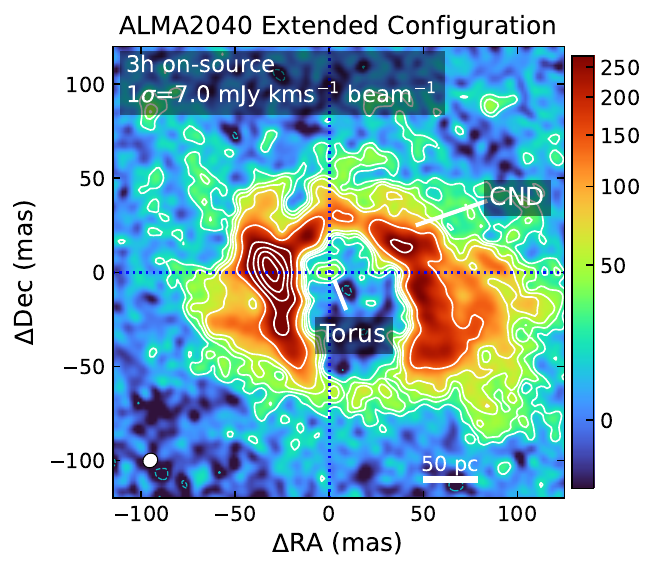}
    \hfill
    \caption{\it Simulated CO $v$=0 (3--2) 345.80\,GHz observations of the central region of a QSO with $L_{\rm bol}$$\sim$10$^{45}$\,erg\,s$^{-1}$ at $z=0.09$, based on ALMA observations of the local Sy2 galaxy NGC~1068 \citep{GarciaBurillo2019A&A...632A..61G}. The location of the torus is indicated by the dotted blue lines.
    The left panel shows a simulated 10\,h on-source integration using the current ALMA extended configuration C-9.
    Although the CND is clearly detected, the limited angular resolution is insufficient to separate the torus and CND emissions. Moreover, current ALMA observations lack the sensitivity to significantly detect the torus emission ($\sim$50\,mJy\,km\,s$^{-1}$\,beam$^{-1}$), even if the torus were spatially resolved from the CND.
    The right panel shows a simulated 3\,h on-source observation using the proposed ALMA2040 Extended Configuration, where the torus is clearly detected and spatially resolved. The integrated-intensity maps are shown in mJy\,km\,s$^{-1}$\,beam$^{-1}$. The contour levels are [3, 5, 7, 10, 20, 40, 60, 80, 100]$\sigma$. The FWHM of the synthesized beam is represented by the white ellipse (19$\times$16\,mas$^2$ for ALMA C-9 and 8$\times$7\,mas$^2$ for ALMA2040 at 317\,GHz).}
    \label{fig:ghost_simulation}
\end{figure*}

\subsection{Torus of high-luminosity AGN - local QSO}

Studies of the resolved torus regions (radii of $\sim$10--30\,pc) in AGN have been restricted to nearby Seyfert-like AGN, as current facilities lack the sensitivity and angular resolution required to study the tori of more luminous AGN, such as QSOs.

Most luminous local Seyferts studied so far present nuclear cold molecular gas deficits with respect to circum-nuclear scales of a few hundreds of parsecs (Fig.~\ref{fig:scales} right). This has been interpreted as an imprint left by AGN feedback when energetic winds clear molecular gas from the torus-scale regions. By contrast, lower-luminosity AGN show higher cold molecular gas concentrations and centrally peaked radial profiles, which is interpreted as these objects being in a build-up phase before the maximum AGN activity \citep{GarciaBurillo2021A&A...652A..98G, GarciaBurillo2024A&A...689A.347G}. One essential piece missing in this scenario is local QSOs. Both type 1 and type 2 QSOs show molecular gas disks and evidence for massive molecular gas outflows on scales of $\sim$kpc \citep[e.g.][]{Molina2021ApJ...908..231M, RamosAlmeida2022A&A...658A.155R}. However, due to their distances $250-500$\,Mpc, it has not been possible to resolve and fully characterize local QSO tori and the SMBH feeding efficiency, as well as to quantify AGN-feedback effects on the molecular gas on nuclear scales.

\subsubsection{Simulations}
In Figure~\ref{fig:ghost_simulation}, we compare the simulated CO $v=0$ (3--2) observation of the torus of a $z$=0.09 QSO by current ALMA in its extended C-9 configuration (left panel) and by ALMA2040 (right panel) using the GHOST simulation tool (\url{https://alma-ghost.jb.man.ac.uk}). 
To simulate the QSO torus, we used existing ALMA CO observations of the nearby (14\,Mpc) Type 2 AGN NGC~1068 \citep{GarciaBurillo2019A&A...632A..61G}. The luminosity was scaled up by a factor of 10, to simulate a typical QSO with  $L_{\rm bol}$$\sim$10$^{45}$\,erg\,s$^{-1}$, $\sim$10 times more luminous than NGC~1068, and the frequency of the line was redshifted to $\sim$317\,GHz with $\sim$20\,km\,s$^{-1}$ channels.
We kept the physical sizes of the torus ($\sim$40\,pc diameter) and circumnuclear disk (CND; $\sim$400\,pc diameter) fixed, scaling their angular sizes to a distance of 420\,Mpc ($z$=0.09).
For current ALMA extended configuration C-9 observations, with a 10\,h on-source integration, the CND is detected. However, the angular resolution is not enough to separate the CND and the torus, and the sensitivity would be insufficient to detect the torus emission, even if it were spatially resolved from the CND.
Instead, with a 3\,h on-source integration using ALMA2040, the torus is detected at $>$7$\sigma$ and is clearly isolated from the CND. Therefore, ALMA2040 will enable accurate measurements of torus masses and sizes in local QSOs, which are not possible with current facilities, but are key to understand the AGN feedback of the dominant population of AGN (i.e., QSO) across cosmic times.

\section{ALMA2040: Technical Requirements}

The proposed science cases require the detection of molecular gas in a wide range of physical conditions, typically in dust-obscured regions, and at high spatial resolution (7--10\,mas for local Seyferts and QSOs). Near- and mid-IR molecular tracers are not ideal, as they are affected by dust extinction and can only trace the warm\slash hot molecular phase.
The 100--350\,GHz range, on the other hand, provides multiple rotational transitions of CO, HCN, and HCO$^+$, which probe from cold and dense to warm molecular gas, and vibrationally excited CO lines, sensitive to the potential hot molecular gas in the dense BLR$^0$. 

Addressing these key questions requires a next-generation sub-mm interferometer with the following technical requirements: 5$\times$ increase in instantaneous spectral line sensitivity at $\sim$300\,GHz, compared to current ALMA ($\sim$4$\times$ compared to anticipated ALMA-WSU). A $\sim$10$\times$ smaller PSF area through $\sim$3$\times$ longer baselines than ALMA-WSU, providing ELT-matched resolution. These can be achieved primarily by increasing the collecting area compared to ALMA-WSU and providing exceptional image fidelity (through exquisite uv-coverage) across a range of angular scales.

\section{Synergies with current and future facilities}
The science cases above resolve nuclear structures and kinematics at a resolution comparable to the Extremely Large Telescope (ELT). 
High-resolution integral-field spectroscopy with HARMONI and imaging with METIS will enable the isolation of the dusty torus emission and help disentangle its contribution from that arising in the surrounding CND. Particularly from warm and hot dust and, where present, polycyclic aromatic hydrocarbon (PAH) features associated with nuclear and/or circumnuclear star-forming activity. 
In addition, these observations will enable the spatially resolved characterization of the ionized gas kinematics in the nuclear regions on scales matched to the ALMA2040 torus molecular maps.
This will enable the measurement of the ionized wind energetics in the same aperture as the molecular torus and BLR$^0$, so that feedback can be characterized on the same scale as feeding. 

In addition, the Square Kilometre Array (SKA) will provide a complementary probe of the neutral atomic phase of AGN feeding and feedback through the \ion{H}{i} 21\,cm line. While ALMA2040 will resolve the molecular component of the torus and circumnuclear disk through CO, HCN, HCO$^+$, [\ion{C}{i}] and vibrationally excited CO transitions, SKA observations will trace the neutral atomic gas associated with AGN-driven outflows and jet-ISM interactions. In particular, \ion{H}{i} absorption against sufficiently bright and compact radio-continuum emission, like that associated with the AGN core and radio jet knots, provides a velocity-resolved probe of cold neutral gas along the line of sight, complementing molecular emission measurements and helping to distinguish inflowing, outflowing and dynamically disturbed components (e.g., \citealt{Pandey-Pommier2026}).

Together with the companion ALMA2040 case for direct dynamical BH mass measurements (Decarli et al. 2026), these synergies with ELT and SKA exploit a key strength of ALMA2040: resolving the cold gas on the scales where BH growth, gas inflows, and AGN feedback are physically connected.

\section*{Acknowledgements}
We thank the anonymous referees for their helpful comments and suggestions, which have improved this manuscript.
This work was developed as part of the broader ALMA2040 community effort. We acknowledge the Lorentz Center and NOVA for supporting the community workshops and coordination activities that helped initiate and shape the ALMA2040 White Paper process.
MPS acknowledges support from grants RYC2021-033094-I and PID2023-146667NB-I00 funded by MCIN/AEI/10.13039/501100011033.
AAH acknowledges support from grant PID2025-168017NB-C22.
\\

\bibliographystyle{apj_mod}

\begin{thebibliography}{41}
\expandafter\ifx\csname natexlab\endcsname\relax\def\natexlab#1{#1}\fi

\bibitem[{{Alonso-Herrero} {et~al.}(2019){Alonso-Herrero},
  {Garc{\'\i}a-Burillo}, {Pereira-Santaella}, {Davies}, {Combes},
  {Vestergaard}, {Raimundo}, {Bunker}, {D{\'\i}az-Santos}, {Gandhi},
  {Garc{\'\i}a-Bernete}, {Hicks}, {H{\"o}nig}, {Hunt}, {Imanishi}, {Izumi},
  {Levenson}, {Maciejewski}, {Packham}, {Ramos Almeida}, {Ricci}, {Rigopoulou},
  {Roche}, {Rosario}, {Schartmann}, {Usero}, \&
  {Ward}}]{AlonsoHerrero2019A&A...628A..65A}
\textbf{Alonso-Herrero} {et~al.} 2019, \aap, 628, A65.
\bibitem[{{Alonso Herrero} {et~al.}(2023){Alonso Herrero},
  {Garc{\'\i}a-Burillo}, {Pereira-Santaella}, {Shimizu}, {Combes}, {Hicks},
  {Davies}, {Ramos Almeida}, {Garc{\'\i}a-Bernete}, {H{\"o}nig}, {Levenson},
  {Packham}, {Bellocchi}, {Hunt}, {Imanishi}, {Ricci}, \&
  {Roche}}]{AlonsoHerrero2023A&A...675A..88A}
\textbf{Alonso Herrero} {et~al.} 2023, \aap, 675, A88.
\bibitem[{{Alonso-Herrero} {et~al.}(2018){Alonso-Herrero}, {Pereira-Santaella},
  {Garc{\'\i}a-Burillo}, {Davies}, {Combes}, {Asmus}, {Bunker},
  {D{\'\i}az-Santos}, {Gandhi}, {Gonz{\'a}lez-Mart{\'\i}n},
  {Hern{\'a}n-Caballero}, {Hicks}, {H{\"o}nig}, {Labiano}, {Levenson},
  {Packham}, {Ramos Almeida}, {Ricci}, {Rigopoulou}, {Rosario}, {Sani}, \&
  {Ward}}]{AlonsoHerrero2018ApJ...859..144A}
\textbf{Alonso-Herrero} {et~al.} 2018, \apj, 859, 144.
\bibitem[{{Antonucci}(1993)}]{Antonucci1993ARA&A..31..473A}
\textbf{Antonucci}, R. 1993, \araa, 31, 473.
\bibitem[{{Audibert} {et~al.}(2019){Audibert}, {Combes}, {Garc{\'\i}a-Burillo},
  {Hunt}, {Eckart}, {Aalto}, {Casasola}, {Boone}, {Krips}, {Viti}, {Muller},
  {Dasyra}, {van der Werf}, \& {Mart{\'\i}n}}]{Audibert2019A&A...632A..33A}
\textbf{Audibert} {et~al.} 2019, \aap, 632, A33.
\bibitem[{{Bisbas} {et~al.}(2015){Bisbas}, {Papadopoulos}, \&
  {Viti}}]{Bisbas2015ApJ...803...37B}
\textbf{Bisbas} {et~al.} 2015, \apj, 803, 37.
\bibitem[{{Bisbas} {et~al.}(2017){Bisbas}, {van Dishoeck}, {Papadopoulos},
  {Sz{\H{u}}cs}, {Bialy}, \& {Zhang}}]{Bisbas2017ApJ...839...90B}
\textbf{Bisbas} {et~al.} 2017, \apj, 839, 90.
\bibitem[{{Combes} {et~al.}(2026){Combes}, {Audibert}, {Garcia-Burillo},
  {Hunt}, {Aalto}, {Casasola}, {Dasyra}, {Eckart}, {Krips}, {Martin}, {Muller},
  {Sakamoto}, {van der Werf}, \& {Viti}}]{Combes2026A&A...705A.124C}
\textbf{Combes} {et~al.} 2026, \aap, 705, A124.
\bibitem[{{Combes} {et~al.}(2019){Combes}, {Garc{\'\i}a-Burillo}, {Audibert},
  {Hunt}, {Eckart}, {Aalto}, {Casasola}, {Boone}, {Krips}, {Viti}, {Sakamoto},
  {Muller}, {Dasyra}, {van der Werf}, \& {Martin}}]{Combes2019A&A...623A..79C}
\textbf{Combes} {et~al.} 2019, \aap, 623, A79.
\bibitem[{{Event Horizon Telescope
  Collaboration}(2019)}]{EHT2019ApJ...875L...1E}
\textbf{Event Horizon Telescope Collaboration}. 2019, \apjl, 875, L1.
\bibitem[{{Event Horizon Telescope Collaboration} {et~al.}(2022){Event Horizon
  Telescope Collaboration}, {Akiyama}, {Alberdi}, {Alef}, {Algaba}, {Anantua},
  {Asada}, {Azulay}, {Bach}, {Baczko}, {Ball}, {Balokovi{\'c}}, {Barrett},
  {Baub{\"o}ck}, {Benson}, {Bintley}, {Blackburn}, {Blundell}, {Bouman},
  {Bower}, {Boyce}, {Bremer}, {Brinkerink}, {Brissenden}, {Britzen},
  {Broderick}, {Broguiere}, {Bronzwaer}, {Bustamante}, {Byun}, {Carlstrom},
  {Ceccobello}, {Chael}, {Chan}, {Chatterjee}, {Chatterjee}, {Chen}, {Chen},
  {Cheng}, {Cho}, {Christian}, {Conroy}, {Conway}, {Cordes}, {Crawford},
  {Crew}, {Cruz-Osorio}, {Cui}, {Davelaar}, {De Laurentis}, {Deane}, {Dempsey},
  {Desvignes}, {Dexter}, {Dhruv}, {Doeleman}, {Dougal}, {Dzib}, {Eatough},
  {Emami}, {Falcke}, {Farah}, {Fish}, {Fomalont}, {Ford}, {Fraga-Encinas},
  {Freeman}, {Friberg}, {Fromm}, {Fuentes}, {Galison}, {Gammie}, {Garc{\'\i}a},
  {Gentaz}, {Georgiev}, {Goddi}, {Gold}, {G{\'o}mez-Ruiz}, {G{\'o}mez}, {Gu},
  {Gurwell}, {Hada}, {Haggard}, {Haworth}, {Hecht}, {Hesper}, {Heumann}, {Ho},
  {Ho}, {Honma}, {Huang}, {Huang}, {Hughes}, {Ikeda}, {Impellizzeri}, {Inoue},
  {Issaoun}, {James}, {Jannuzi}, {Janssen}, {Jeter}, {Jiang},
  {Jim{\'e}nez-Rosales}, {Johnson}, {Jorstad}, {Joshi}, {Jung}, {Karami},
  {Karuppusamy}, {Kawashima}, {Keating}, {Kettenis}, {Kim}, {Kim}, {Kim},
  {Kim}, {Kino}, {Koay}, {Kocherlakota}, {Kofuji}, {Koch}, {Koyama}, {Kramer},
  {Kramer}, {Krichbaum}, {Kuo}, {La Bella}, {Lauer}, {Lee}, {Lee}, {Leung},
  {Levis}, {Li}, {Lico}, {Lindahl}, {Lindqvist}, {Lisakov}, {Liu}, {Liu},
  {Liuzzo}, {Lo}, {Lobanov}, {Loinard}, {Lonsdale}, {Lu}, {Mao}, {Marchili},
  {Markoff}, {Marrone}, {Marscher}, {Mart{\'\i}-Vidal}, {Matsushita},
  {Matthews}, {Medeiros}, {Menten}, {Michalik}, {Mizuno}, {Mizuno}, {Moran},
  {Moriyama}, {Moscibrodzka}, {M{\"u}ller}, {Mus}, {Musoke}, {Myserlis},
  {Nadolski}, {Nagai}, {Nagar}, {Nakamura}, {Narayan}, {Narayanan},
  {Natarajan}, {Nathanail}, {Fuentes}, {Neilsen}, {Neri}, {Ni}, {Noutsos},
  {Nowak}, {Oh}, {Okino}, {Olivares}, {Ortiz-Le{\'o}n}, {Oyama}, {{\"O}zel},
  {Palumbo}, {Paraschos}, {Park}, {Parsons}, {Patel}, {Pen}, {Pesce},
  {Pi{\'e}tu}, {Plambeck}, {PopStefanija}, {Porth}, {P{\"o}tzl}, {Prather},
  {Preciado-L{\'o}pez}, \& {Psaltis}}]{EHT2022ApJ...930L..12E}
\textbf{Event Horizon Telescope Collaboration} {et~al.} 2022, \apjl, 930, L12.
\bibitem[{{Gallimore} {et~al.}(2016){Gallimore}, {Elitzur}, {Maiolino},
  {Marconi}, {O'Dea}, {Lutz}, {Baum}, {Nikutta}, {Impellizzeri}, {Davies},
  {Kimball}, \& {Sani}}]{Gallimore2016ApJ...829L...7G}
\textbf{Gallimore} {et~al.} 2016, \apjl, 829, L7.
\bibitem[{{Garc{\'\i}a-Burillo} {et~al.}(2021){Garc{\'\i}a-Burillo},
  {Alonso-Herrero}, {Ramos Almeida}, {Gonz{\'a}lez-Mart{\'\i}n}, {Combes},
  {Usero}, {H{\"o}nig}, {Querejeta}, {Hicks}, {Hunt}, {Rosario}, {Davies},
  {Boorman}, {Bunker}, {Burtscher}, {Colina}, {D{\'\i}az-Santos}, {Gandhi},
  {Garc{\'\i}a-Bernete}, {Garc{\'\i}a-Lorenzo}, {Ichikawa}, {Imanishi},
  {Izumi}, {Labiano}, {Levenson}, {L{\'o}pez-Rodr{\'\i}guez}, {Packham},
  {Pereira-Santaella}, {Ricci}, {Rigopoulou}, {Rouan}, {Shimizu}, {Stalevski},
  {Wada}, \& {Williamson}}]{GarciaBurillo2021A&A...652A..98G}
\textbf{Garc{\'\i}a-Burillo} {et~al.} 2021, \aap, 652, A98.
\bibitem[{{Garc{\'\i}a-Burillo} {et~al.}(2019){Garc{\'\i}a-Burillo}, {Combes},
  {Ramos Almeida}, {Usero}, {Alonso-Herrero}, {Hunt}, {Rouan}, {Aalto},
  {Querejeta}, {Viti}, {van der Werf}, {Vives-Arias}, {Fuente}, {Colina},
  {Mart{\'\i}n-Pintado}, {Henkel}, {Mart{\'\i}n}, {Krips}, {Gratadour}, {Neri},
  \& {Tacconi}}]{GarciaBurillo2019A&A...632A..61G}
\textbf{Garc{\'\i}a-Burillo} {et~al.} 2019, \aap, 632, A61.
\bibitem[{{Garc{\'\i}a-Burillo} {et~al.}(2016){Garc{\'\i}a-Burillo}, {Combes},
  {Ramos Almeida}, {Usero}, {Krips}, {Alonso-Herrero}, {Aalto}, {Casasola},
  {Hunt}, {Mart{\'\i}n}, {Viti}, {Colina}, {Costagliola}, {Eckart}, {Fuente},
  {Henkel}, {M{\'a}rquez}, {Neri}, {Schinnerer}, {Tacconi}, \& {van der
  Werf}}]{GarciaBurillo2016ApJ...823L..12G}
\textbf{Garc{\'\i}a-Burillo} {et~al.} 2016, \apjl, 823, L12.
\bibitem[{{Garc{\'\i}a-Burillo} {et~al.}(2024){Garc{\'\i}a-Burillo}, {Hicks},
  {Alonso-Herrero}, {Pereira-Santaella}, {Usero}, {Querejeta},
  {Gonz{\'a}lez-Mart{\'\i}n}, {Delaney}, {Ramos Almeida}, {Combes},
  {Angl{\'e}s-Alc{\'a}zar}, {Audibert}, {Bellocchi}, {Davies}, {Davis},
  {Elford}, {Garc{\'\i}a-Bernete}, {H{\"o}nig}, {Labiano}, {Leist}, {Levenson},
  {L{\'o}pez-Rodr{\'\i}guez}, {Mercedes-Feliz}, {Packham}, {Ricci}, {Rosario},
  {Shimizu}, {Stalevski}, \& {Zhang}}]{GarciaBurillo2024A&A...689A.347G}
\textbf{Garc{\'\i}a-Burillo} {et~al.} 2024, \aap, 689, A347.
\bibitem[{{Gendriesch} {et~al.}(2009){Gendriesch}, {Lewen}, {Klapper},
  {Menten}, {Winnewisser}, {Coxon}, \&
  {M{\"u}ller}}]{Gendriesch2009A&A...497..927G}
\textbf{Gendriesch} {et~al.} 2009, \aap, 497, 927.
\bibitem[{{Gravity Collaboration} {et~al.}(2018){Gravity Collaboration},
  {Sturm}, {Dexter}, {Pfuhl}, {Stock}, {Davies}, {Lutz}, {Cl{\'e}net},
  {Eckart}, {Eisenhauer}, {Genzel}, {Gratadour}, {H{\"o}nig}, {Kishimoto},
  {Lacour}, {Millour}, {Netzer}, {Perrin}, {Peterson}, {Petrucci}, {Rouan},
  {Waisberg}, {Woillez}, {Amorim}, {Brandner}, {F{\"o}rster Schreiber},
  {Garcia}, {Gillessen}, {Ott}, {Paumard}, {Perraut}, {Scheithauer},
  {Straubmeier}, {Tacconi}, \& {Widmann}}]{Gravity2018Natur.563..657G}
\textbf{Gravity Collaboration} {et~al.} 2018, \nat, 563, 657.
\bibitem[{{Greenhill} {et~al.}(1996){Greenhill}, {Gwinn}, {Antonucci}, \&
  {Barvainis}}]{Greenhill1996ApJ...472L..21G}
\textbf{Greenhill} {et~al.} 1996, \apjl, 472, L21.
\bibitem[{{Greenhill} {et~al.}(1995){Greenhill}, {Henkel}, {Becker}, {Wilson},
  \& {Wouterloot}}]{Greenhill1995A&A...304...21G}
\textbf{Greenhill} {et~al.} 1995, \aap, 304, 21.
\bibitem[{{Harrison} \& {Ramos Almeida}(2024)}]{Harrison2024Galax..12...17H}
\textbf{Harrison} {et~al.} 2024, Galaxies, 12, 17.
\bibitem[{{Imanishi} {et~al.}(2018){Imanishi}, {Nakanishi}, {Izumi}, {Wada},
  {Wada}, {Wada}, \& {Wada}}]{Imanishi2018ApJ...853L..25I}
\textbf{Imanishi} {et~al.} 2018, \apjl, 853, L25.
\bibitem[{{Imanishi} {et~al.}(2020){Imanishi}, {Nguyen}, {Wada}, {Hagiwara},
  {Iguchi}, {Izumi}, {Kawakatu}, {Nakanishi}, \&
  {Onishi}}]{Imanishi2020ApJ...902...99I}
\textbf{Imanishi} {et~al.} 2020, \apj, 902, 99.
\bibitem[{{Impellizzeri} {et~al.}(2019){Impellizzeri}, {Gallimore}, {Baum},
  {Elitzur}, {Davies}, {Lutz}, {Maiolino}, {Marconi}, {Nikutta}, {O'Dea}, \&
  {Sani}}]{Impellizzeri2019ApJ...884L..28I}
\textbf{Impellizzeri} {et~al.} 2019, \apjl, 884, L28.
\bibitem[{{Khouri} {et~al.}(2016){Khouri}, {Vlemmings}, {Ramstedt}, {Lombaert},
  {Maercker}, \& {De Beck}}]{Khouri2016MNRAS.463L..74K}
\textbf{Khouri} {et~al.} 2016, \mnras, 463, L74.
\bibitem[{{Kormendy} \& {Ho}(2013)}]{Kormendy2013ARA&A..51..511K}
\textbf{Kormendy} {et~al.} 2013, \araa, 51, 511.
\bibitem[{{Krolik} \& {Begelman}(1988)}]{Krolik1988ApJ...329..702K}
\textbf{Krolik} {et~al.} 1988, \apj, 329, 702.
\bibitem[{{Lo}(2005)}]{Lo2005ARA&A..43..625L}
\textbf{Lo}, K.~Y. 2005, \araa, 43, 625.
\bibitem[{{Maloney} {et~al.}(1996){Maloney}, {Hollenbach}, \&
  {Tielens}}]{Maloney1996ApJ...466..561M}
\textbf{Maloney} {et~al.} 1996, \apj, 466, 561.
\bibitem[{{Mercedes-Feliz} {et~al.}(2026){Mercedes-Feliz},
  {Angl{\'e}s-Alc{\'a}zar}, {Cevallos}, {Oh}, {Garc{\'\i}a-Burillo},
  {Cochrane}, {Ramos Almeida}, {Faucher-Gigu{\`e}re}, {Alonso-Herrero},
  {Richings}, {Pereira-Santaella}, {Moreno}, {Roy}, {D{\'\i}az-Santos}, \&
  {Hopkins}}]{MercedesFeliz2026}
\textbf{Mercedes-Feliz} {et~al.} 2026, arXiv e-prints, arXiv:2606.25025.
\bibitem[{{Mercedes-Feliz} {et~al.}(2023){Mercedes-Feliz},
  {Angl{\'e}s-Alc{\'a}zar}, {Hayward}, {Cochrane}, {Terrazas}, {Wellons},
  {Richings}, {Faucher-Gigu{\`e}re}, {Moreno}, {Su}, {Hopkins}, {Quataert}, \&
  {Kere{\v{s}}}}]{MercedesFeliz2023}
\textbf{Mercedes-Feliz} {et~al.} 2023, \mnras, 524, 3446.
\bibitem[{{Molina} {et~al.}(2021){Molina}, {Wang}, {Shangguan}, {Ho}, {Bauer},
  {Treister}, \& {Shao}}]{Molina2021ApJ...908..231M}
\textbf{Molina} {et~al.} 2021, \apj, 908, 231.
\bibitem[{{Ochmann} {et~al.}(2024){Ochmann}, {Kollatschny}, {Probst},
  {Romero-Colmenero}, {Buckley}, {Chelouche}, {Chini}, {Grupe}, {Haas},
  {Kaspi}, {Komossa}, {Parker}, {Santos-Lleo}, {Schartel}, \&
  {Famula}}]{Ochmann2024A&A...686A..17O}
\textbf{Ochmann} {et~al.} 2024, \aap, 686, A17.
\bibitem[{{Pandey-Pommier} {et~al.}(2026){Pandey-Pommier}, {Sethuram}, {Konar},
  {Muthumariappan}, {Roy}, {Roy}, \& {Marcowith}}]{Pandey-Pommier2026}
\textbf{Pandey-Pommier} {et~al.} 2026, arXiv e-prints, arXiv:2608.19162.
\bibitem[{{Papadopoulos} {et~al.}(2018){Papadopoulos}, {Bisbas}, \&
  {Zhang}}]{Papadopoulos2018MNRAS.478.1716P}
\textbf{Papadopoulos} {et~al.} 2018, \mnras, 478, 1716.
\bibitem[{{Ramos Almeida} {et~al.}(2022){Ramos Almeida}, {Bischetti},
  {Garc{\'\i}a-Burillo}, {Alonso-Herrero}, {Audibert}, {Cicone}, {Feruglio},
  {Tadhunter}, {Pierce}, {Pereira-Santaella}, \&
  {Bessiere}}]{RamosAlmeida2022A&A...658A.155R}
\textbf{Ramos Almeida} {et~al.} 2022, \aap, 658, A155.
\bibitem[{{Springel} {et~al.}(2005){Springel}, {Di Matteo}, {Hernquist}, {al.},
  {al.}, {al.}, {al.}, \& {al.}}]{Springel2005MNRAS.361..776S}
\textbf{Springel} {et~al.} 2005, \mnras, 361, 776.
\bibitem[{{Thi} \& {Papadopoulos}(2024)}]{Thi2024A&A...688L..20T}
\textbf{Thi} {et~al.} 2024, \aap, 688, L20.
\bibitem[{{Urry} \& {Padovani}(1995)}]{Urry1995PASP..107..803U}
\textbf{Urry} {et~al.} 1995, \pasp, 107, 803.
\bibitem[{{Uttley} {et~al.}(2014){Uttley}, {Cackett}, {Fabian}, {Kara}, \&
  {Wilkins}}]{Uttley2014A&ARv..22...72U}
\textbf{Uttley} {et~al.} 2014, \aapr, 22, 72.
\bibitem[{{Vogelsberger} {et~al.}(2020){Vogelsberger}, {Marinacci}, {Torrey},
  \& {Puchwein}}]{Vogelsberger2020}
\textbf{Vogelsberger} {et~al.} 2020, Nature Reviews Physics, 2, 42.
\end{thebibliography}

\end{document}